# Rapid Classification of Glaucomatous Fundus Images

HARDIT SINGH,[1*] SIMARJEET SAINI,[2,] AND VASUDEVAN LAKSHMINARAYANAN[2,3]

[1]Cameron Heights Collegiate Institute, 301 Charles St E, Kitchener, ON, N2G 2P8, Canada
[2] Department of Electrical and Computer Engineering, University of Waterloo, 200, University Ave. W., Waterloo, ON, N2L 3G1, Canada
[3]School of Optometry and Vision Science, University of Waterloo, 200, University Ave. W., Waterloo, ON, N2L 3G1, Canada.
*hsharditsingh@gmail.com

**Abstract:** We propose a new method for training convolutional neural networks which integrates reinforcement learning along with supervised learning and use it for transfer learning for classification of glaucoma from colored fundus images. The training method uses hill climbing techniques via two different climber types viz. "random movement" and "random detection" integrated with supervised learning model through stochastic gradient descent with momentum (SGDM) model. The model was trained and tested using the Drishti-GS, and RIM-ONE-r2 datasets having glaucomatous and normal fundus images. The performance for prediction was tested by transfer learning on five CNN architectures namely, GoogLeNet, DenseNet-201, NASNet, VGG-19, and Inception-Resnet-v2. A 5-fold classification was used for evaluating the performance and high sensitivities while maintaining high accuracies were achieved. Of the models tested, DenseNet-201 architecture performed the best in terms of sensitivity and area under the curve (AUC). This method of training allows transfer learning on small datasets and can be applied for tele-ophthalmology applications including training with local datasets.

## 1. Introduction

Glaucoma is the "silent killer" of eyesight and is the second leading cause for blindness [1]. An estimated 64 million people suffer from glaucoma worldwide. If left undiagnosed, it causes irreversible damage to the optic nerve which eventually leads to blindness. Presently, there is no cure, and treatment only slows the progression of the disease. Hence, it is imperative that glaucoma is diagnosed at an early stage. Unfortunately, eye care is not integrated with the universal health care system in most countries and it is estimated that over 70% of cases are left undiagnosed in early stages [2]. The prevalence of glaucoma increases with age and the population of the elderly is expected to double in the next decade putting a strain on an already stretched health care systems [3]. Thus, a paradigm shift in glaucoma detection is required where patients can be screened rapidly and accurately, all at a low cost. Detection of glaucoma typically relies on examination of structural damage to the optic nerve combined with measurements of visual function such as visual fields and intraocular pressure. A review of glaucoma diagnostic methods is given in the paper by Sharma et al., (2008) [4].

Currently, there are three common tests for diagnosis of glaucoma. These include the contact or non-contact tonometry to measure the intraocular pressure (IOP) in the eye, evaluation of the visual field, and evaluation of the optic nerve head (ONH) using optical means such as ophthalmoscopy, and/or fundus imaging [5]. IOP measurements are not enough on their own as many patients do not show signs of elevated IOP. Further, in many cases by the time there is an appreciable IOP increase, neural damage has already occurred. The visual field only gets affected at advanced stages of the disease and thus, again misses the early stages of the disease. Imaging technologies for evaluating the ONH are only available in tertiary health centers and patients are only referred to them once the vision has been affected. Complete diagnosis requires further tests including gonioscopy (a measurement of the "angle" of theanterior chamber of the eye), and pachymetry, which measures the thickness of the cornea [5].

The complete set of tests are time-consuming and expensive and requires a glaucoma specialist. Thus, ophthalmologists are looking for accurate screening methods. Evaluation of the ONH is considered to have the most promise for effective screening [6]. Figure 1 shows the ONH region cropped from a fundus image. Some of the features which are used to screen for glaucoma are also indicated in the figure. The ONH consists of the optic disc and the optic cup. The optic disc is the point for the exit of the ganglion cell axons which make up the optic nerve. The optic cup area is the brightest part of the optic disc. Evaluation of the cup to disc ratio is used to detect glaucoma as the reduction of the optic nerve fibers creates optic disc cupping and thus, increasing the ratio of the cup diameter to the disc diameter. Typically cup to disc ratio of a healthy eye is smaller than 0.5. If the ratio is greater than 0.7, it is considered to be an indicator of severe glaucoma [7]. However, the cup to disc ratio is not always correct as this cupping can also be hereditary and would remain stable with time. Also, these measurements are made more difficult because the disc can be tilted. Thus, diagnosis is made by evaluating the progression of the cup to disc ratio (CDR) over several months or even years. There is also the so-called ISNT rule. The ONH region is divided into different regions called the inferior (I) at the south pole of the ONH, superior (S) at the north pole of the ONH, nasal (N) on the inside of the eye facing the nose and temporal (T) on the outside of the eye. The thickness of the neuroretinal rim which is the space between the optic disc edge and optic cup edge is measured and for a normal eye follows the pattern called the ISNT rule. The rule states that in normal eyes, the thickness of the neuroretinal rim for the $I \geq S \geq N \geq T$. On its own, the ISNT rule is also not accurate enough. In a study conducted by Harizman et al., (2006) on 109 subjects, the ISNT rule was intact in 79% of normal eyes and 28 % of glaucoma eyes [8]. Thus, 28 % of people who had glaucoma would be incorrectly diagnosed as normal creating false negatives. A recent study showed that it was more accurate to measure the neuroretinal rim to optic disk ratio (RDR) as compared to CDR or ISNT rule [9]. Classification accuracy of 84.4, 75.9, and 92.2 % was achieved using CDR, ISNT, and RDR respectively.

Currently, in practice, ophthalmologists manually segment different parts of the image and use their experience to diagnose looking at different features. The quality of the diagnosis is only as good as the judgment of the doctor. In a study done by Almazroa et al., (2017) [10], it was found that the accuracy for segmenting the optic disc and optic cup varied between 75% to 90% for 6 different ophthalmologists. The ophthalmologists only agreed with each other in approximately 75% of the cases. Ophthalmologists also look at other features in the retinal images e.g. how the blood vessels kink in the ONH region. These features are missed when classification is done based only on image segmentation. The problem of classifying images is where computer vision excels and thus, there has been tremendous effort around the world to use computer vision to classify normal and glaucomatous images over the last few years [11, 12]**.** Feature-based classification can be broadly split up into two different categories based on the classification methods used. Features of the image like the pixel intensity histograms [13], wavelets [14], Fourier transforms and spline coefficients, higher order spectra analysis, and texture-based features [15] are extracted from preprocessed images. The identified features are used to train a classifier and make a decision based on the prevalence of specific features. Classification is done either through simple classifiers like support vector machines or random forest classification or through deep learning methods involving convolutional neural networks [16].

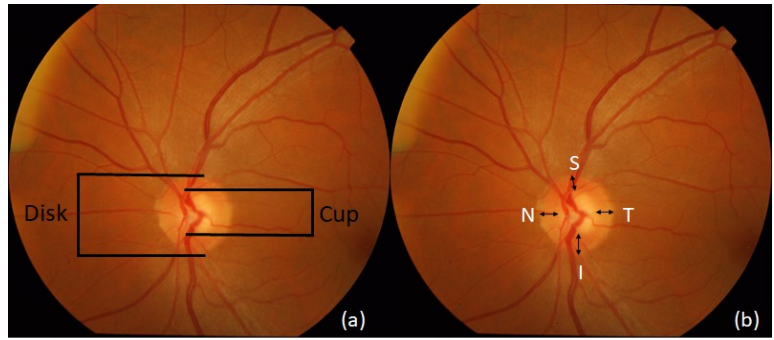

Fig. 1. Examples of features used to detect Glaucoma in a colored fundus image, (a) describes the markings for the cup and the disc which can be used to calculate the cup to disc ratio. (b) marks the inferior(I), superior(S), nasal(N), and temporal(T) thicknesses of the neuroretinal rim which can be used for classifying healthy eyes.

Of the various classification methods, CNN has shown the most promise. In CNN, feature extraction, feature learning, and classification all happen simultaneously. A network or model is created by many layers called neurons which transform an input image to outputs (glaucoma present/absent). There are three major strategies that use CNNs: training and building a network from scratch, using pre-trained CNN features, and conducting unsupervised CNN pre-training with supervised fine-tuning. Table 1 summarizes the work done using CNNs. Chen et al., (2015) [17] used a 6 layer CNN to classify optic disc and optic cup and achieved 83.1 % accuracy using a private dataset. Al-Bander et al., (2017) [18] used a 23 layer CNN architecture to extract the features but did the classification using an SVM algorithm and achieved an accuracy of 88.2%. Fu et al., (2018) [19] created an ensemble of 4 CNN's to create a new architecture called D-Net to achieve an accuracy of 83.2% and 66.6 % on two different data sets, respectively.

**Table 1. CNN models for classification of glaucoma: a comparison**

| Author, Year | Method | Number of Images | Success Rate(Acc) |
|---|---|---|---|
| Al-Bander et al., 2017 [18] | CNN 23 layers and SVM | 255 Normal 200 Glaucoma | Accuracy 88.2% Sensitivity 84.78% |
| Chai et al., 2018 [39] | MB-NN | 2,554 | Accuracy 91.51% Sensitivity 92.33% Specificity 90.90% |
| Chen et al., 2015 [17] | CNN(6 layers) | 482 Normal 168 Glaucoma | AUC 83.1% |
| Christopher et al., 2018 [9] | Transfer Learning with ResNet, VGG16 and Inception-v3 | 9189 Normal 5633 Glaucoma | Sensitivity 88% Specificity 95% |

| Fu et al., 2018 [19] | Ensemble of 4 CNNs | 482 Normal 168 Glaucoma | Sensitivity 84.78% Specificity 83.80% |
|---|---|---|---|
| Li et al., 2018 [20] | Transfer Learning with Inception Network | 48116 | Sensitivity 95.6% Specificity 92.0% |
| Shibata et al., 2018 [41] | Transfer Learning with ResNet | 1768 Normal 1364 Glaucoma | AUC – 96.5% |
| Singh et al., 2019 [21] | Transfer learning with Inception V-3 | 191 images | Accuracy: 92.5 % Sensitivity: 92.3 % Specificity: 93.6 % |

Impressive results have been achieved by Li et al., (2018) [20] using an Inception-v3 CNN architecture [21] on a private database with 48,000+ images to detect glaucomatous images. The images of the dataset were labeled by 21 ophthalmologists. A local space average color substitution was implemented to account for varying levels of illumination. The large dataset with a deep model led to high accuracies approaching 94 %. Singh et al., (2019) [22] showed that transfer based learning can be used to train Inception V3 for small datasets and achieved an accuracy of 92.5 %.

For deep learning methods to be used effectively by ophthalmologists, it is also important to win their confidence. One of the problems is simply this: deep learning methods act as black boxes with limited visibility into decision making and thus, has restricted clinical use. Recent explainability studies aim to show the features that influence the decision of a model the most. The reader is referred to Singh et al., (2021) [23] for a review of the new field of explainable AI. Another major problem is that fundus cameras and lighting conditions vary and to win that confidence, it may be necessary to train the deep learning networks with local datasets that may not have a large number of images. Deep learning algorithms are prone to overfitting the data. In this paper, we propose and implement a new training algorithm combining supervised learning with reinforcement based learning for transfer training of CNN architectures. The aim of this research was to develop a new transfer learning training method that allows for rapid training on CNN architectures. The performance of 5 different architectures viz. DenseNet-201 [24], VGG-19 [25], NASNet [26], Inception-Resnet-v2 [27], GoogLeNet [28] was evaluated using cross-validation for classification of glaucomatous fundus images. The trained models were optimized for sensitivity while maintaining high accuracy.

## 2. Methodology

### 2.1 Dataset Description

Two publically available datasets namely RIM-ONE-r2 and DRISHTI-GS were chosen for this study. The RIM-ONE r2 dataset is a sub-dataset of the Open Retinal Image Database for Optic Nerve Evaluation (RIM-ONE) [29]. The dataset consists of 455 colored fundus images out of which 255 are normal and 200 have glaucoma or suspected to have glaucoma. The dataset has a region of interest, the ONH region, already extracted with different image sizes. DRISHTI-GS [30] is a public dataset that consists of 101 uncropped fundus images. The images have been annotated by 5 clinicians into normal and glaucoma classes. An image is considered to be glaucomatous or not based on the majority decision. Seventy of the images in the dataset are in class glaucoma while 31 images are normal. The images of DRISHTI-GS have a Field-of-View of 30º with an image size of 2896 ×1944 pixels. The patients' age varied from 40 to 80 years with roughly equal numbers of males and females. A critical overview of fundus image databases for deep learning ophthalmic diagnosis is given by Sengupta et al., (2020) [31].

*2.2 Statistical Analysis*

A 5-fold cross-validation was used to train the models and test the performance. The normal and glaucomatous images were first randomly rank ordered and then a Venetian blind was used to create the five folds. Each fold consisted of 204 normal images from the RIM-ONE-r2 dataset, 25 normal images from Drishti-GS (one fold had 24 normal images from Drishti-GS) for a total of 229 normal images, 160 glaucomatous images from the RIM-ONE-r2, and 56 images from DrishtiGS for a total of 216 glaucomatous images. For every fold, the performance was tested on 51 independent normal images and 40 independent glaucomatous images from RIM-ONE-r2 dataset; 6 normal images (one fold had 7 normal images) and 14 glaucomatous images from Drishti-GS. Within an epoch, 80 % of the images used in the training data set were used for training while 20 % were used for validation of the models. The split was done randomly and the images were shuffled at the beginning of each epoch.

*2.3 Pre-Processing*

The only pre-processing used was to format the image to the correct input size for which the CNN architectures were originally trained on. For GoogLeNet, VGG-19, and DenseNet-201, the input size of the image was adjusted to 224 × 224 pixels while for Inception-ResNet-v2, the image size was adjusted to 299 × 299 pixels and for NASNet, the image size was adjusted to 331 × 331 pixels. No algorithms were used to correct the lighting conditions or to increase the contrast.

RIM-ONE-r2 images already have the ONH region cropped and the images were used as is. For the Drishti-GS images, an algorithm was developed for autocropping the retinal images to extract the ONH region. Autocropping was based on the idea that the ONH region is the largest brightest region in the fundus image and the various steps are shown in Figure 2. To successfully target the region, the images were preprocessed to enhance the contrast. First, the red channel of the RGB image was extracted as it showed a clear ONH region compared to other color channels. The extracted red channel image was then contrast-enhanced by scaling the brightest pixel intensity to the maximum value of 255 and scaling the other pixels according to their intensity values. The image was then segmented into 50 "superpixels" using the simple linear iterative clustering (SLIC) algorithm [32]. The algorithm connects pixels with similar intensities together reducing the complexity of segmentation. After superpixelation was done, the intensity values of each superpixel were averaged to create a coarse image that highlights the OD further. The average intensity superpixel image was then thresholded to create a binary image. Superpixels above the threshold were converted to an intensity of 255 and superpixels below the threshold are converted to 0. A threshold of 254 gave the highest accuracy in extracting the region of interest. Once, the image was thresholded, the largest region in the image was isolated and the centroid and area of the region were extracted using morphological operations [33]. The centroid was considered to be the center of the ONH region and a square with the length of pixels desired was used to crop the original image. The method took 7 seconds to crop the image on a Lenovo ThinkPad T480 equipped with Intel I-core 5, $8^{th}$ generation processor, and 8 GB of memory. The accuracies of the ROI extraction are summarized in Table 2. The accuracy of the implemented ROI extraction method was calculated on the Drishti-GS database and Magrabia, and Bin Rushed images of the RIGA dataset [34]. The accuracy was measured by observing whether the ONH region was completely within the square cropped region. Almazroa et al., (2017) [10] had previously used morphological operations and Type-I Fuzzy technique to find the brightest regions for ONH extraction on Magrabia and Bin Rushed images and achieved success rates of 93.6 % and 94.8 % respectively for the two databases. Our method of using superpixels and morphological operations decreased the failures by nearly half while requiring much lower computing resources. Only for two images in the Drishti-GS database, the ROI was not correctly centered. A manual extraction was implemented for these images.

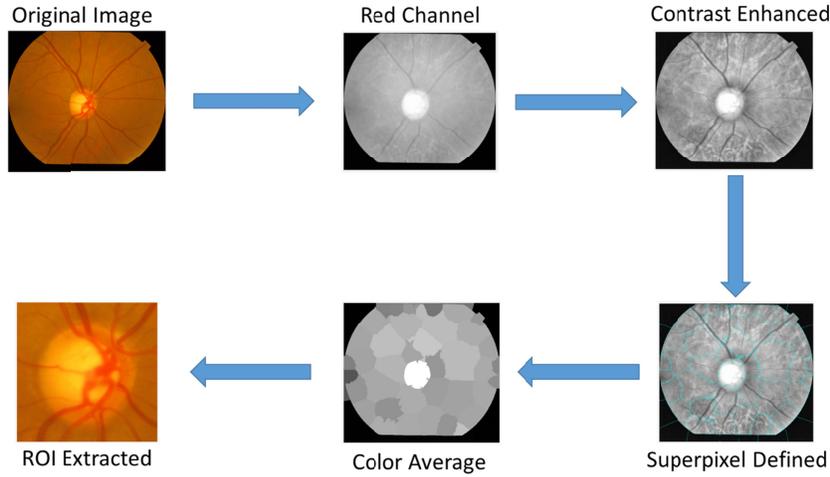

Fig. 2. shows the methods for extracting the optic nerve head image from a full sized color fundus image.

Table 2. Comparison of the results for ROI extraction.

| Dataset | Our Work | Ahmed et al., 2017 [10] |
|---|---|---|
| Drishti | 98.0 % | N.A. |
| Magrabia | 97.9 % | 93.6 % |
| Bin Rushed | 97.4 % | 94.8 % |

*2.4 Data Augmentation*

Since CNNs are deep learning architectures, there is a major concern of overfitting the data. To prevent this, data augmentation was implemented. The following operations were applied randomly on the training set within the specified ranges – rotation -10 to + 10 degrees; shear -0.2 to 0.2; zoom up to 10%; horizontal flip; width and height shift up to 10 %. At the start of each epoch, a combination of these operations was randomly applied to an image in the training set to create partially overlapping views of the data during the training. Further, patches of the required image size were randomly generated during each iteration.

*2.5 Transfer Learning Training Method*

Normally in the training phase for transfer learning, supervised learning is used where the network model is given training images and accuracy of classification measured as the weights of the deep layers are changed. The weights are changed based on optimization functions e.g., Adam optimizer [35]. However, this can lead to overfitting in small size datasets. A new way of training was developed which integrated reinforcement learning along with supervised learning. The Training dataset was grouped into mini-batches of size 64 and in an epoch, the complete dataset was covered. Multiple hill climbers were created and randomly assigned a "mode". One of the climbers was assigned to optimize using stochastic gradient descent with momentum (SGDM)) while other climbers were randomly assigned one of the two modes viz. "random movement" and "random detection". When the "random detection" mode was assigned, climbers try to repeat the same movement as their last step. If a climber did not make progress that is larger than model variation strength, it would randomly activate one of its detectors to find a better path. Once one of the detectors found slopes where climbers can improve in a value that is larger than model variation strength, the climbers would stop activating the remaining detectors and start to move in that direction. As for the mode "random movement", climbers just move completely randomly in trying to optimize the weights of the network. At the end of the epoch, the climber which achieves the highest validation accuracy survives and new climbers are created and randomly assigned to one of the modes. The process continues until the validation accuracy starts to decrease at which time the training stops or maximum number of epochs are achieved. Table 3 summarizes the hyper-parameters including the maximum number of epochs, initial learning rate for SGDM mode, and the number of iterations in an epoch for transfer learning on the different CNN architectures.

Table 3. Training settings used for the transfer learning process.

| Starting Network | Learning Rate | No. of Epochs | No. of iterations per Epoch, n |
|---|---|---|---|
| GoogLeNet | $1\times10^{-3}$ | 60 | 6 |
| Inception-Resnet v2 | $7\times10^{-4}$ | 100 | 6 |
| VGG-19 | $1\times10^{-4}$ | 30 | 6 |
| DenseNet-201 | $1\times10^{-3}$ | 30 | 6 |
| NASNet | $1\times10^{-3}$ | 20 | 22 |

*2.6 Performance Metrics*

To evaluate the performance of the trained models, the Receiver Operating Characteristics (ROC) were created and the Area under the Curve (AUC) was utilized. The AUC does not depend upon the classification threshold and thus, can be used to compare different models. Also, three other criteria were used for evaluating the performance, namely, accuracy, sensitivity, and specificity and we also looked at the four outcomes for an image that is classified viz. true positive (TP), false negative (FN), true negative (TN) and false negative (FN). Since our goal was to create rapid screening algorithms, it is important that no glaucomatous image is missed as the cost to the patient is very high. Thus, a higher emphasis was put on the sensitivity. Obviously, specificity should also be considered as a large number of false positives is not optimal.

*2.7 Freezing Initial Layers*

It has been previously suggested that retraining all the layers is more effective than freezing the initial layers [22] though no comparisons were made. This is because the fundus images are quite different as compared to the images used in the ImageNet dataset that the CNN architectures have been previously trained on. We compared the results by freezing the first 10 layers with retraining all the layers for one fold of one of NASNet and found that the results were better by freezing the first 10 layers. As such, for the extensive studies done, the first 10 layers were frozen for all networks.

## 3. Results and Discussion

### 3.1 Freezing Layers

The CNN networks being used for training were originally built to classify objects quite different from color fundus images. The low level and mid-level features of the images could be very different. An initial experiment was run to see whether to freezing the first 10 layers while training or allowing training to change the weights in all layers. The experiment was run by training NASNet network and testing on one fold to see which method achieved the highest accuracy. NASNet was chosen for the network as it is the deepest network of those studied. Table 4 summarizes the TP, TN, FN, and FP for the two experiments. Training done with freezing first 10 layers achieved higher accuracy than having to optimize the weights in all the layers. The first 10 layers only look into very low-level features in the image and while the images are quite different for glaucoma, the low-level features are probably not that different. These networks had originally been trained with over 1 million images and thus, the robustness of the weights for the first 10 layers is very strong. By choosing to train them over only ~ 450 images, we may have not reached optimal values.

Table 4. Comparison of the results when first 10 layers are frozen versus optimizing the weights of all the layers. The experiment was done on a single fold by transfer learning on NASNet architecture.

|  | Freezing weights of first 10 layers | Changing weights of all layers |
| --- | --- | --- |
| True Positive | 35 | 34 |
| False Negative | 5 | 6 |
| True Negative | 50 | 47 |
| False Positive | 1 | 4 |

### 3.2 Comparison of CNN Architectures for RIM-ONE-r2 database

Figure 3 shows the confusion matrices for the 5 networks studied. The NASNet achieved the highest accuracy and specificity and did equally well in diagnosing glaucoma and normal cases. DenseNet-201 achieved the highest sensitivity only classifying 7 glaucomatous images incorrectly. The sensitivity and specificity for DenseNet-201 was 96.5% and 91.0% respectively and thus, and only has a 3.5% chance of missing a case. The VGG-19 architecture had consistent results in both sensitivity and specificity but performed worse than NASNet and DenseNet-201 in overall accuracy making it an unviable architecture. In previous work, VGG-19 had shown the best performance among CNN architectures but NASNet and DenseNet-201 were not studied [36]. GoogLeNet and Inception-Resnet-v2 performed similarly, both achieving a sensitivity of 89.5% and specificities of around 91%. Although the GoogLeNet architecture achieved lower accuracies than the previously mentioned networks, it should be noted that all the images were classified in rapid times (<0.5 seconds) and thus, it may be suitable for mobile platforms.

An important parameter to judge the network's ability to accurately classify images is to investigate the results achieved over different folds. A stable and accurate network should have consistent performance in different folds. Figure 4 plots the accuracy, sensitivity, and specificity for the different networks over the 5 independent folds and the values along the standard deviations are summarized in Table 5. If we look at the overall accuracy over the different folds, DenseNet201, NASNet and VGG-19 had consistent performance over the folds with standard deviations (SD) of 1.43 %, 2.20 %, and 1.74 % respectively. The SD for GoogLeNet and Inception-Resnet v2 was 4.10 % and 3.84 %. However, when we also consider the sensitivity and specificity, DenseNet-201 has the most consistent performance and achieved values greater than 90 % sensitivity for all folds. In some folds, NASNet achieved 100 % sensitivity but has lower specificity for those same folds. Interestingly, different networks performed relatively differently in different folds demonstrating that these networks were focusing on different features of the images. GoogLeNet had the highest variation showing that the training was not stable over the network.

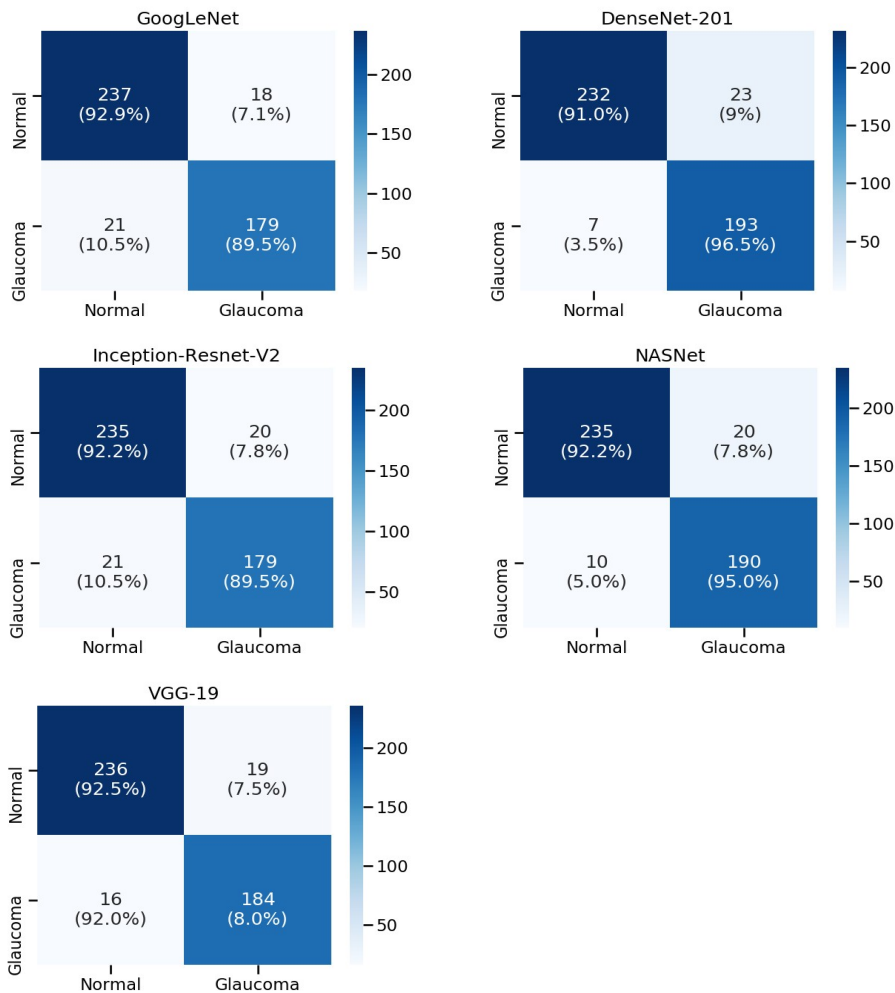

Fig. 3. Confusion matrices for the test data set for different trained CNN architectures. Results from the test samples from five folds were combined.

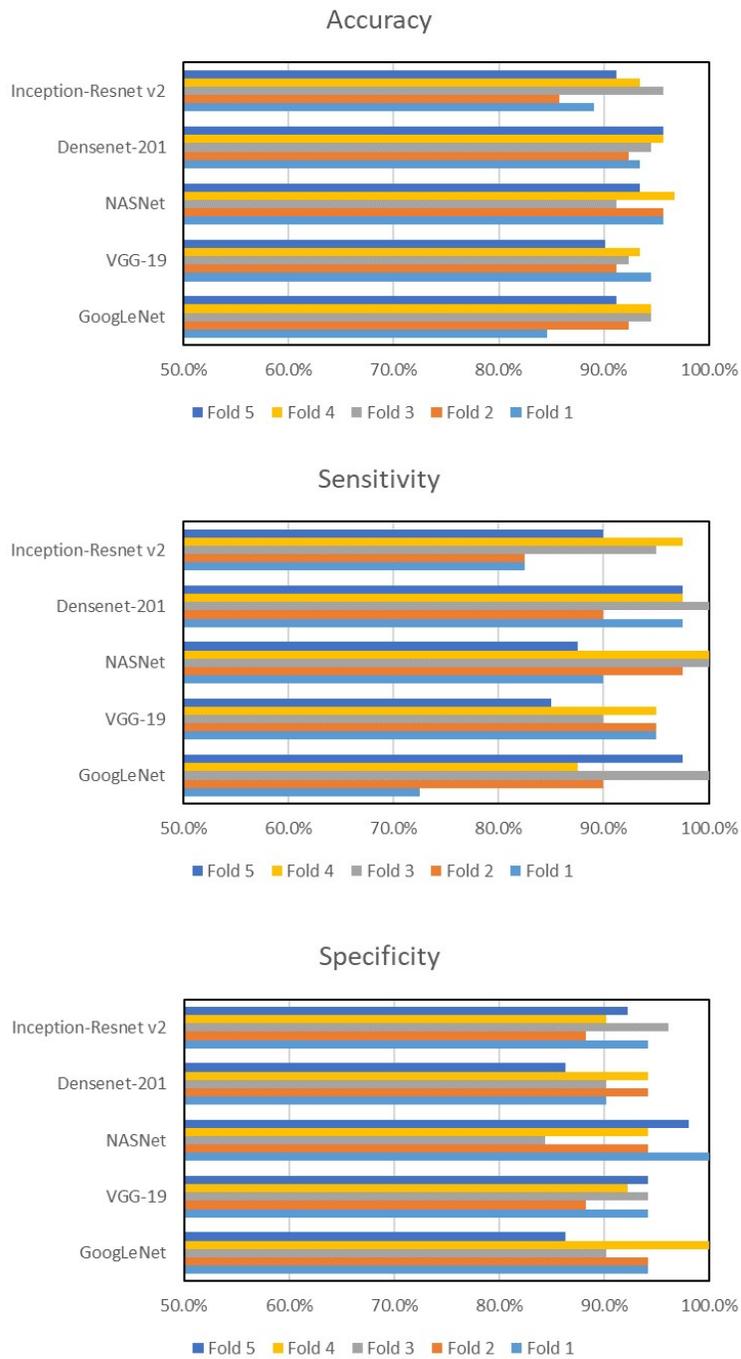

Figure 4. Comparison of (a) Accuracy; (b) Sensitivity; and (c) Specificity for 5 different folds of each network. DenseNet-201 and VGG-19 had the most consistent results over the five folds.

Table 5. Comparison of the accuracy, sensitivity and specificity for the CNN architectures. SD values are the standard deviations over the 5 different folds. AUC = Area under the curve

| Network | Accuracy | | Sensitivity | | Specificity | | AUC |
|---|---|---|---|---|---|---|---|
| | Overall (%) | SD (%) | Overall (%) | SD (%) | Overall (%) | SD (%) | |
| **Tested on RIM-ONE-r2 dataset** | | | | | | | |
| GoogLeNet | 91.4 | 4.07 | 89.5 | 10.81 | 92.9 | 5.11 | 0.970 |
| Inception-Resnet v2 | 91.0 | 3.84 | 89.5 | 6.94 | 92.2 | 3.10 | 0.973 |
| VGG-19 | 92.3 | 1.74 | 92.0 | 4.47 | 92.5 | 2.56 | 0.983 |
| DenseNet-201 | 93.4 | 1.43 | 96.5 | 3.79 | 91.0 | 3.28 | 0.990 |
| NASNet | 94.5 | 2.20 | 95.0 | 5.86 | 94.1 | 6.04 | 0.978 |
| **Tested on Drishti-GS dataset** | | | | | | | |
| DenseNet-201 | 90.1 | 3.54 | 98.6 | 3.19 | 71.0 | 7.22 | 0.91 |
| VGG-19 | 86.1 | 2.18 | 97.1 | 3.91 | 61.3 | 7.60 | 0.87 |
| NASNet | 89.1 | 2.30 | 100 | 0 | 64.5 | 8.25 | 0.90 |

ROC curves are plotted in Figure 5 (a) for the 5 different architectures studied and the AUC values are also summarized in Table 5. All the networks had an AUC over 97 % with DenseNet-201 performing the best with AUC of 0.990. NASNet which achieved the highest accuracy among the networks had a lower AUC of 0.977. The ROC curves show that all the networks classified the images robustly. Considering all the performance metrics, DenseNet-201 has the best performance among the networks studied. For the ImageNet dataset, DenseNet-201 performs worse than VGG19 and nearly the same as Inception-Resnet v2. The results here show that the accuracy on the ImageNet dataset may not correspond to the same accuracy when the networks are trained for other applications and hence, one needs to consider all networks for transfer training. For the DenseNet201 network, if the threshold for classification is reduced to 0.4 instead of 0.5, then 98 % of glaucoma images are classified correctly while the accuracy only drops to 93.2 % from 93.5 % (there is only one extra false positive).

To understand the robustness and stability of the DenseNet-201 network ROC curves were analyzed for each of the five folds. A stable network should have consistent results over all the folds. Figure 5 (b) plots the ROC curves for each fold along with the ROC curve for all the images. The AUC varied from 0.980 to 0.998 with a standard deviation of 0.007.

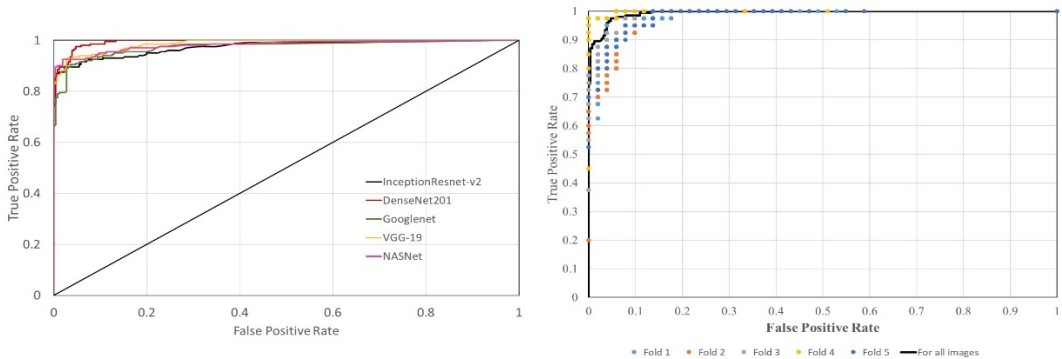

Figure 5 (a) ROC curved for the 5 different CNN architectures studied. (b) ROC curved for 5 different folds for DenseNet-201 network.

*3.3 Comparison of CNN Architectures for Drishti-GS database*

For Drishti-GS, the top performing models viz. DenseNet-201, VGG-19, and NASNet were tested and the confusion matrices are shown in Figure 6. The results are also summarized in Table 5. DenseNet-201 achieved the highest overall accuracy of 89.1 % but misclassified one glaucomatous image as normal. NASNet classified all glaucomatous images correctly but misclassified two normal images more than DenseNet-201. While the sensitivity was high for the three networks, the specificity was much lower. The images have large light intensity variations and reflection artifacts in them. Further, the 5 ophthalmologists all agree only on 52 images, roughly half of the dataset. These Fifty-two images were correctly classified by the DenseNet-201 and NASNet and the corresponding number for VGG-19 was 51.

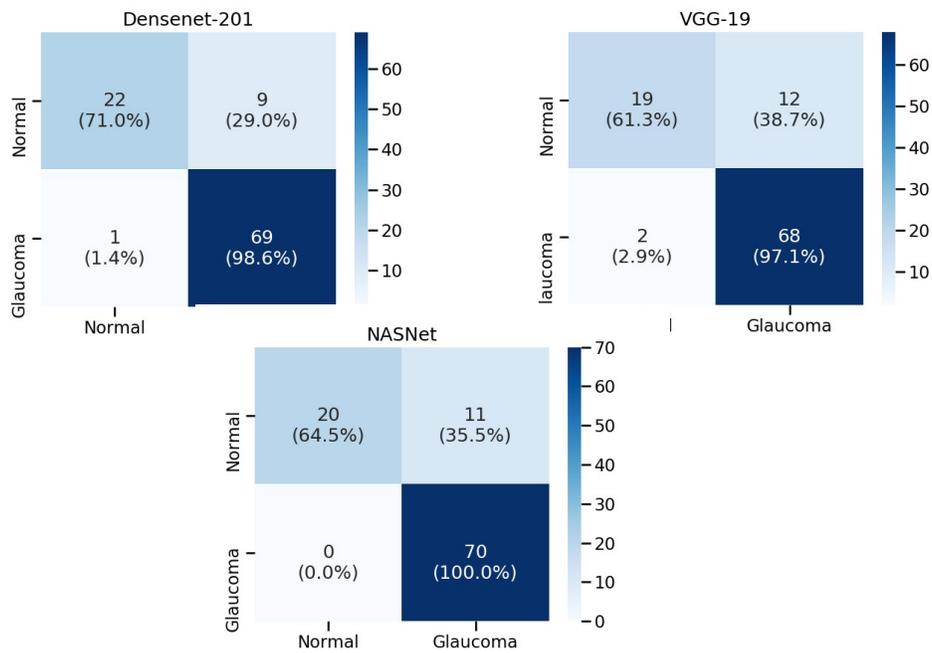

Fig 6. Confusion matrices for Drishti-GS for DenseNet-201, VGG-19, and NASNet.

The class decision on the dataset was made as a majority decision by 5 ophthalmologists with varying experience from 3-20 years. We recall Richard Feynman's, (1988) [37] caveat about taking averages with people of varying expertise in his analysis of the space shuttle Challenger crash investigation. When results from people with different experiences are averaged, the average brings down the accuracy, not increases it. If the criterion is changed such that the image is glaucomatous if only two experts classify it as glaucoma, then the classification results improve and are summarized in Table 6. DenseNet-201 outperformed the other networks and achieved an accuracy of 93.1 %, sensitivity of 97.3 %, and specificity of 80.8 %.

Table 6. Comparison of the accuracy, sensitivity and specificity for Drishti-GS if an image is classified as glaucoma provided two of the 5 ophthalmologists classify it as glaucoma.

|  | NASnet | VGG-19 | DenseNet201 |
|---|---|---|---|
| **True Positive** | 72 | 71 | 73 |
| **False Negative** | 3 | 4 | 2 |
| **True Negative** | 17 | 17 | 21 |
| **False Positive** | 9 | 9 | 5 |
|  |  |  |  |
| **Accuracy** | 88.1% | 87.1% | 93.1% |
| **Sensitivity** | 96.0% | 84.7% | 97.3% |
| **Specificity** | 65.4% | 65.4% | 80.8% |

*3.5 Comparison with Other Works*

Table 7 summarizes some of the pertinent results of our work and compares it with other CNN classifications for glaucoma. As can be seen from the table, the results we have achieved are better than that reported so far both in terms of accuracy and sensitivity. The difference is a lot more significant when you compare methods using the same database. For example, results that were achieved with VGG-19 are much better than those achieved by Gomez-Valverde et al., (2019) [36] and Diaz-Pinto et al., (2019) [38].

Table 7. Comparison of our work with previous studies using CNN architectures.

| Author Name | Method | Database | No. of images tested | Accuracy | Sensitivity | Specificity |
|---|---|---|---|---|---|---|
| **Our work** | DenseNet | RIM-ONE-r2 | 455 | 93.4 % (93.2 %) | 96.5 % (98.0 %) | 91.0% (90.6 %) |
| **Our work** | DenseNet | Drishti | 101 | 90.1 % | 98.6 % | 71.0 % |
| **Our work** | VGG-19 | RIM-ONE-r2 | 455 | 92.3 % | 92.0 % | 93.5% |
| **Our work** | GoogLeNet | RIM-ONE-r2 | 455 | 91.4 % | 89.5 % | 92.9 % |
| **Singh et al. [22]** | Inception v3 | RIM-ONE | 187 | 92.5 % | 92.3 % | 93.5 % |
| **Gomez-Valverde et** | VGG-19 | RIM-ONE r2 | 191 | 88.5 % | 88.6 % | 88.4 % |
| al. [36] | | | | | | |
| Gomez-Valverde et al. [36] | VGG-19 | Drishti | 17 | 88.2 % | 91.7 % | 80.0 % |
| Li et al. [21] | Inception v3 | Private Database | 48116 | - | 95.6 % | 92.0 % |
| Chai et al. [39] | MB-neural network | Private Database | 2554 | 91.5 % | 92.3 % | 90.9 % |
| Cerentini et al. [41] | GoogLeNet | Drishti | 101 | 86.2 % | - | - |
| Diaz-Pinto et al. [37] | VGG-19 | ACRIMA + Drishti + RIMONE+ shchoi86-HRF + HRF | - | 90.7 % | 92.4 % | 88.5 % |

## 4. Conclusion and Discussion

A new method was developed for transfer learning by combining supervised learning with reinforcement learning. Different CNN networks such as GoogLeNet, VGG-19, DenseNet-201, NASNet, and Inception-ResNet-v2 have been used to classify glaucomatous images. These networks were used in this paper. GoogLeNet was designed for smartphone applications and is a very quick network to train. VGG-19 has previously shown the highest accuracy for the RIM-ONE r2 dataset [30]. NASNet is really deep and has been shown to be accurate for the Imagenet dataset. DenseNet-201 has similar accuracies to VGG-19 though at a slower classification time. However, this network has never been tried before. Inception-Resnet v2 is Resnet with added inception layers and gives higher accuracies than VGG-19 for the Imagenet dataset. The proposed algorithm is tested on publically available dataset and the results show that the model can be used to effectively train and deploy CNN architectures for glaucoma diagnosis. The method was tested for using 5-fold classification for glaucomatous images and achieved very high sensitivities while maintaining high accuracies. Transfer learning on DenseNet-201 performed the best achieving an AUC of 0.99, and sensitivity of 96.5 % for the RIMONE-r2 dataset. If the threshold is changed to 0.4, then the sensitivity increases to 98 %. For Drishti-GS dataset, DenseNet-201 achieved an accuracy of 90.1 %, AUC of 0.91 and sensitivity of 98.6 %. The results achieved in this work are better than what has been reported so far for the same datasets before, demonstrating that the new training method improves the accuracy of prediction. In the future, the authors plan to deploy the DenseNet-201 model into the medical field for real world testing. This will allow to see what effects the model takes when tested blindly on new images. The authors also plan to train the model on images with different stages of glaucoma as this can indicate the correct treatment a glaucomatous patient needs.

**Acknowledgments**

VL is also with the departments of Physics and Systems Design Engineering at UW and acknowledges support by a Discovery grant from NSERC.

**Disclosures**

The authors declare no conflicts of interest.